\documentclass[conference,11pt]{IEEEtran}
\usepackage{graphicx}
\usepackage{amsmath}
\usepackage{amssymb}
\usepackage{graphics}
\usepackage{float}
\usepackage{epstopdf}
\usepackage{psfrag}

\begin{document}

%
\title{Two-way Wireless Video Communication using Randomized Cooperation, Network Coding and Packet Level FEC }


\author{\IEEEauthorblockN{Xiaozhong Xu, \"{O}zg\"{u}~Alay, Elza Erkip, Yao Wang and Shivendra Panwar}
\IEEEauthorblockA{Department of Electrical and Computer Engineering\\
Polytechnic Institute of New York University, Brooklyn, NY, USA\\
\{xiaozhongxu, ozgualay\}@gmail.com, \{elza, yao\}@poly.edu, panwar@catt.poly.edu}
}


%


\maketitle

\begin{abstract}
Two-way real-time video communication in wireless networks requires high bandwidth, low delay and error resiliency. This paper addresses these demands by proposing a system with the integration of Network Coding (NC), user cooperation using Randomized Distributed Space-time Coding (R-DSTC) and packet level Forward Error Correction (FEC) under a one-way delay constraint. Simulation results show that the proposed scheme significantly outperforms both conventional direct transmission as well as R-DSTC based two-way cooperative transmission, and is most effective when the distance between the users is large. 

\end{abstract}


%
\IEEEpeerreviewmaketitle

\let\thefootnote\relax\footnote{This work is supported in part by NSF Grant CNS-0905446, by the New York State Center for Advanced Technology in Telecommunications (CATT), and by the Wireless Internet Center for Advanced Technology (WICAT), an NSF Industry/University
Cooperative Research Center.}
\vspace*{-12pt}
\section{Introduction}\label{sec:introduction}
Recently, with the enrichment of mobile terminal devices, real-time applications such as mobile video conferencing have become popular. The design of two-way communication systems for these applications requires a robust and low delay solution that can maintain high data rates. One approach is to allow two users exchange their data with the help of relays. An efficient way of using relays in two-way communication systems is network coding (NC), where the data sent by relays is a function of the data received from both users. The most commonly used NC algorithm sends the difference of two incoming packets from two users by using bit-wise ``XOR" \cite{XOR_08}\cite{OCNC_08}\cite{Cui_DSTC09}.

Simultaneous transmission by multiple relays is considered to be an efficient approach of user cooperation, and can be achieved using a distributed space-time coding (DSTC) \cite{STC_99}. The basic idea is to coordinate and synchronize the relays such that each relay acts as one antenna of a regular STC. However, DSTC places a high requirement on feedback and synchronization among nodes, which is a challenge in dynamic wireless environments where potential relays change over time. Scenarios using DSTC with a fixed number of relays are studied together with NC in \cite{OCNC_08}\cite{Cui_DSTC09}. To reduce the control overhead of DSTC, Randomized DSTC (R-DSTC) \cite{RDSTC_07} provides a decentralized solution where a variable number of relays are used. In R-DSTC, relays do not need to be aware of the presence (or absence) of other relays and decide autonomously whether to forward or not. Furthermore, they can send packets simultaneously without the need for tight synchronization \cite{RDSTC_09}.

In the presence of packet losses, end-to-end feedback and retransmission can be introduced to recover possible errors \cite{OCNC_08}. However, the use of retransmission is usually avoided or limited for two-way real-time communication due to the stringent delay requirements. Instead, Forward Error Correction (FEC), which requires no feedback from the receiver, can be adopted.

In this paper we will address a two-way cooperative communication system design with delay constraint. As opposed to the approach in \cite{OCNC_08}, we consider a low protocol overhead approach. We integrate R-DSTC \cite{alay_RDSTC10}, network coding and packet level FEC \cite{alay_monet09} to provide a robust and efficient two-way system with low delay. We analyze different proposed cooperative schemes in an IEEE 802.11g \cite{IEEE802.11} based network and compare their performances with direct transmission.

The rest of this paper is organized as follows: Section II presents the proposed system; Section III gives the problem formulation and optimization; in Section IV simulation results are shown and discussed; Conclusions are addressed in Section V.

\section{System Model}\label{sec:sysmodel}
We study a wireless network where two users exchange video sequences with each other. We assume these users are among a set of randomly distributed nodes, all of which are equipped with one antenna and can transmit at different transmission rates supported by the underlying physical layer. We assume the channel between each pair of nodes experience path loss and independent fading that is constant over the transmission time of a single packet. Also each node is only able to send or receive (but not both) at the same time.

\begin{figure}
	\centerline{\includegraphics[width=0.40\textwidth]{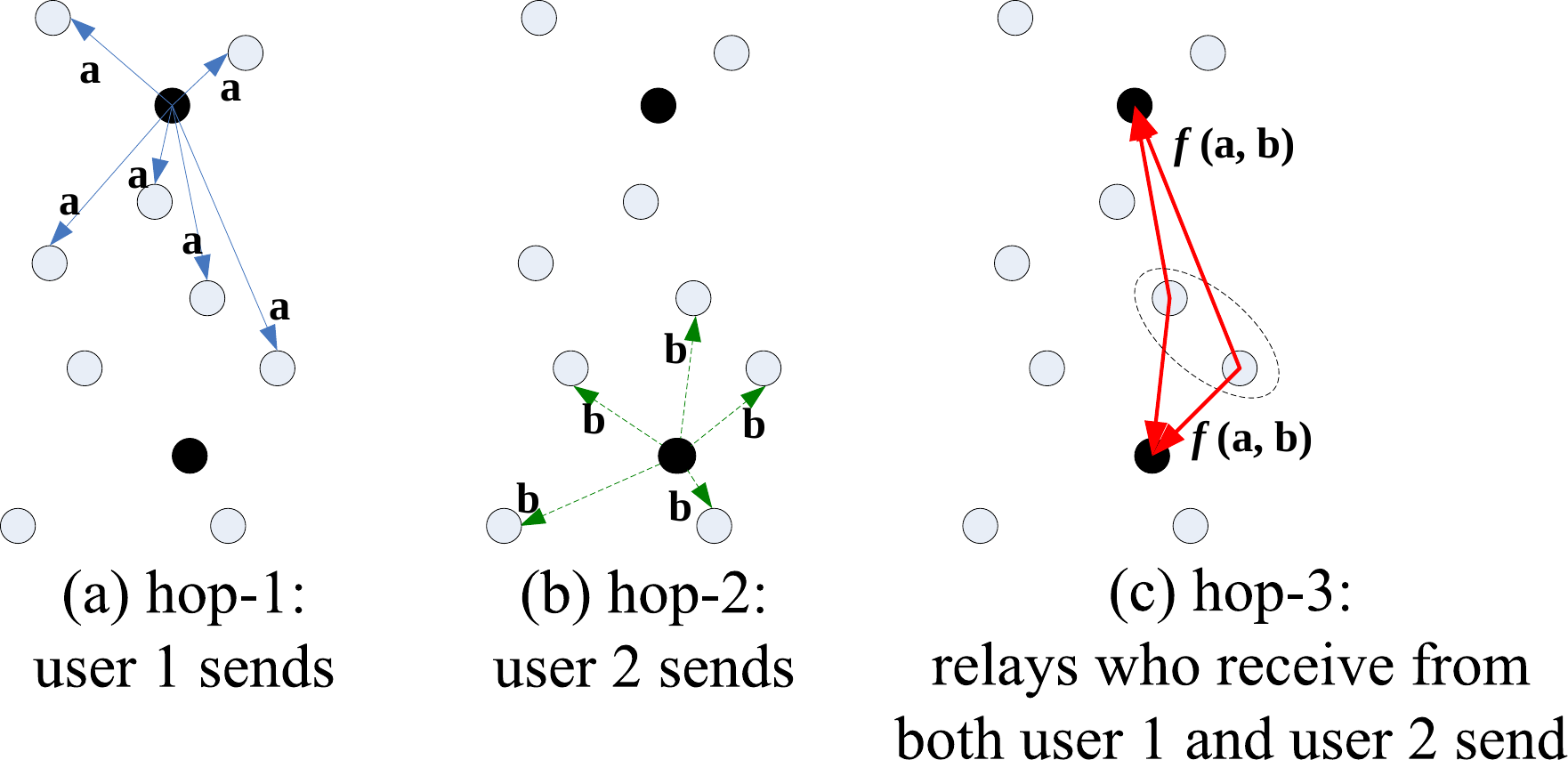}}
	\vspace{-5pt}
	\caption{
		Two-way communication using R-DSTC+NC
	}
	\vspace{-20pt}
	\label{fig: RDSTC+NC}
\end{figure}

In our system (see Fig. \ref{fig: RDSTC+NC}), after each user transmits a packet to the other user, the relays  that can decode both source packets use NC to combine (XOR) the two packets and transmit the combination to both users at the same time. The XORed packet are sent using R-DSTC. This is done by passing the data in the XORed packet through an STC encoder at each relay. The output of the STC encoder is in the form of {\it L} parallel streams with each stream corresponding to an antenna in an {\it L}-antenna system. Each relay transmits a random linear weighted combination of all {\it L} streams. Each receiver estimates the equivalent channel gain using pilot signals, and decodes the R-DSTC signal using a conventionally designed decoder for STC reception. On receiving from relays, each user is able to decode the packet sent by the other user by making use of its own. Or, each user can also decode direct transmission from the other user.

To recover lost packets due to uncorrectable errors at the physical layer, each source estimates the end-to-end packet loss rate, including the effect of the relays, and uses packet level FEC (Reed-Soloman code) to append necessary parity packets when transmitting.

For the baseline direct transmission, we assume user 1 knows the average channel quality (in terms of the average received SNR) between user 2 and itself. User 1 transmits packets to user 2 at rate $R_{12}$ and FEC rate $\gamma_{12}$ under a delay constraint. The FEC rate $\gamma_{12}$ is chosen based on the channel packet error rate (PER) $p_{12}$ such that user 2 receives the packets with an FEC decoding failure rate less than $\tau$. Similar considerations also apply to user 2. We refer this system as ``Direct Transmission".

We will also consider another baseline system that uses the relays, but for sequential transmission of each user's packets, as opposed to NC. In this case, user 1 transmits packets to user 2 at rate $R_{12}$. For each packet, a node that receives the packet correctly becomes a relay. Note that the set of relays can be different from one packet to another. All such nodes re-encode and relay the packet simultaneously to user 2 at a rate of $R_{r2}$ using R-DSTC. User 2 also attempts to decode packets sent from user 1 directly. The FEC rate here is determined based on the end-to-end PER from user 1 to user 2 such that user 2 decodes all the source packets (from the received source and parity packets, from either user 1 directly or through the relays) with an FEC decoding failure rate less than $\tau$. Similar considerations apply to that from user 2 to user 1. We refer this system as ``R-DSTC".

For the proposed R-DSTC+NC approach, we use the common relays that receive both users' packets. In this case user 1 transmits a packet at rate $R_{12}$ to user 2 (decoded by some of the nodes, denoted as $S_{1}$) and user 2 transmits a packet at rate $R_{21}$ to user 1 (decoded by nodes $S_{2}$). Those nodes in $S=S_{1}\cap S_{2}$ form the relay set and send the XOR of the two received packets at a rate of $R_{r}$ to both users using R-DSTC. Each user can correctly recover the other user's packet if it receives the packet directly from the other user or the XOR packet from the relay set. Similarly, we determine the end-to-end FEC rate from user 1 to user 2 such that user 2 receives the packets with an FEC failure rate less than $\tau$. This system will be called as ``R-DSTC+NC".   


\section{Optimization of Transmission Parameters}\label{sec:formula}

There are various system parameters that must be determined, including the transmission rates at each hop, the FEC rate at the source,  FEC block size, and the transmission time of each hop. As mentioned earlier, the FEC rate is determined so that the end-to-end FEC decoding failure rate is below a threshold $\tau$. We choose $\tau$ to be sufficiently low so that the packet loss effect in the decoded video is negligible. Under this strategy, video quality is directly proportional to the useful video rate from user 1 to user 2, $R_{v,12}$ and from user 2 to user 1, $R_{v,21}$. We further require that the time duration of video data (corresponding to a certain number of video frames) that form an FEC block from each user to be under a delay threshold $T_{delay}$, which affects the total one-way delay. This is because a receiver can decode an FEC block only after all video data inside this block becomes available at the sender (which takes $T_{delay}$ time), processed by the video encoder and then the FEC encoder, and finally transmitted. $T_{delay}$ is the minimal one-way delay from video capture at the sender to video display at the receiver. In practice, video encoding, FEC encoding, transmission, FEC decoding and finally video decoding will add additional delay. Considering that maximum one way delay that is acceptable for video call applications is typically below 350 ms \cite{350}, $T_{delay}$ is set to 150 ms in this work.


We determine optimum operating parameters (e.g. $R_{12}, R_{21}, R_{r2}, R_{r1}, R_r$) to maximize the average video rate, $(R_{v,12}+ R_{v,21})/2$, under the delay constraint. While deriving the video rate, we assume that only a portion of the wireless transmission air time is allocated for the two-way video call, and denote this portion (called effective data ratio) by $\beta$. We further assume the packet size is a constant, denoted by $P_{size}$ bits. We next derive the relation between the achievable video rates and the operating parameters of different schemes in Section \ref{sec:sysmodel}, and describe how to optimize the operating parameters.
\subsection{Scheme I: Direct Transmission}

For the direct system, we consider a rate adaptive direct transmission scheme \cite{alay_monet09} under the delay constraint.  Note that video packets in one FEC block from both senders must be delivered within $T_{delay}$ time duration, to sustain the real-time two-way communications. Because we only use $\beta$ portion of the air time for the video call, the available transmission time is further reduced to $\beta T_{delay}$. The total transmission time is divided into two phases. In the first phase (using $t_{I}$ portion of the total time $\beta T_{delay}$), user 1 transmits its data (both source and parity packets) to user 2; similarly in the second phase (using $t_{II}=1-t_{I}$ portion of $\beta T_{delay}$) user 2 transmits its data to user 1. The total number of packets that can be sent  by user 1 with a transmission rate of $R_{12}$ is:
\begin{equation}
 n_{12}=\lfloor \beta R_{12} t_{I} T_{delay} / P_{size} \rfloor
\end{equation}
Note that since $\beta$, $T_{delay}$ and $P_{size}$ are fixed, $n_{12}$ is a function of $R_{12}$ and $t_{I}$. Among the $n_{12}$ packets, we must determine the number of source packets $s_{12}$ and the number of parity packets $m_{12}$, so that $s_{12}+m_{12}=n_{12}$. The ratio $s_{12}/n_{12}$ is the FEC rate. We assume the use of a perfect code (such as Reed-Solomon (RS) code) so that we can recover all the $s_{12}$ source packets as long as the number of lost packets is at most $m_{12}$.

Let $p_{12}$ denote the average packet error rate (PER) from user 1 to user 2. It depends on the distance between two users, and transmission rate $R_{12}$, and can be determined by channel simulations. For a given $n_{12}$, and assuming the packet loss events are i.i.d., the probability of receiving at least $s_{12}$ packets out of $n_{12}$ packets at user 2 can be formulated as:
\vspace{-2pt}	
\begin{equation}
P(s_{12})= \sum_{s=s_{12}}^{n_{12}} {{n_{12}}\choose{s}} {p_{12}}^{(n_{12}-s)}{(1-p_{12})}^s   \label{prob_single}
\end{equation}
\vspace{-5pt}	

The FEC decoding failure rate from user 1 to user 2 is $1-P(s_{12})$. We find through numerical search the maximum $s_{12}$, such that $(1-P(s_{12}))$ $\leq$ $\tau$. Because  $p_{12}$ is a function of $R_{12}$, $n_{12}$ is a function of $R_{12}$ and $t_{I}$, $s_{12}$ is also a function of $R_{12}$ and $t_{I}$, denoted by $s_{12} (R_{12},t_{I})$.

After the computation of $s_{12}$, the video rate from user 1 to user 2 can be expressed as:
\vspace{-2pt}	
\begin{equation}
 R_{v,12}= s_{12}(R_{12},t_{I}) P_{size} / T_{delay} \label{RV_12}
\end{equation}
\vspace{-2pt}	
Similarly, the video rate from user 2 to user 1 is:
\vspace{-2pt}	
\begin{equation}
 R_{v,21}= s_{21}(R_{21},t_{I}) P_{size} / T_{delay}  \label{RV_21}
\end{equation}
For a given distance between the two users, we search over all sustainable ($R_{12}$, $R_{21}$, $t_{I}$) through an exhaustive search to choose the optimum set that maximizes the average video rate. Note that the candidate rates and time division in all simulations in this paper are from a discrete space, with the rates chosen from those provided in the IEEE 802.11g standard.

\subsection{Scheme II: R-DSTC}
For the system using R-DSTC, we similarly divide $T_{delay}$ into two phases. In phase I, using $t_{I}$ portion of $T_{delay}$, user 1 transmits packets (both source and parity) to user 2,  using two hops transmission for each packet. In the first hop, user 1 sends the packet using rate $R_{12}$. In the second hop, all nodes receiving this packet relay it to user 2 using R-DSTC at rate $R_{r2}$.  The total time needed to transmit one packet is $T=P_{size}/R_{12}+P_{size}/R_{r2}$. Therefore the total number of packets that can be sent over the duration $\beta t_{I} T_{delay}$ is:
\vspace{-2pt}	
\begin{equation}
n_{12} = \lfloor  \beta t_{I} T_{delay} / T \rfloor =  \Big\lfloor \frac{\beta t_{I} T_{delay} R_{12} R_{r2}}{ P_{size} (R_{12} +R_{r2})} \Big\rfloor \label{eq_RDSTC}
\end{equation}
\vspace{-10pt}

Similarly as in Scheme I, we can use \eqref{prob_single} to get the maximum $s_{12}$ given $n_{12}$. However, the $p_{12}$ in \eqref{prob_single} now depends on $R_{12}$, $R_{r2}$, the user distance and node distribution. This is because node distribution affects who can act as relays for given $R_{12}$, and it also affects whether user 2 can receive from these relays, for given $R_{r2}$. For a given node placement, we determine $p_{12}$ from each pair of candidate $R_{12}$ and $R_{r2}$ through channel simulations. The useful video rates can be found using equations similar as \eqref{RV_12} and \eqref{RV_21}. We search over all candidate $R_{12}$, $R_{r2}$, $R_{21}$, $R_{r1}$, $t_{I}$ to maximize the average video rate.

\subsection{Scheme III: R-DSTC+NC}

For the system using R-DSTC and NC, we let user 1 send a packet at rate $R_{12}$, user 2 send a packet at rate $R_{21}$, and those nodes who receive both packets XOR the two packets and send the resulting packet using R-DSTC at rate $R_r$. The total time needed to transmit one packet from each sender is $T=P_{size}( 1/R_{12} + 1/R_{21}  + 1/R_{r})$. The total number of packets  that can be delivered by each sender over a duration of $\beta T_{delay}$ is:

\vspace{-17pt}	

\begin{eqnarray}
 \lefteqn{n_{12}=n_{21} = \lfloor \beta T_{delay} /T\rfloor}  \nonumber \\
 && =\Big\lfloor \frac{ \beta T_{delay}R_{12} R_{21} R_r} { P_{size} (R_{21} R_r + R_{12} R_r + R_{12} R_{21})} \Big\rfloor \label{eq_NC}
\end{eqnarray}


Given $R_{12}$, $R_{21}$, $R_r$, and the node placement, one can find the end-to-end PER, $p_{12}$ (from user 1 to user 2) and $p_{21}$, which are different in general. Then (2) is used to determine the number of source packets $s_{12}$ and $s_{21}$, for given $\tau$. 
We note that although the total numbers of packets in each direction must be the same, the numbers of source packets can differ. For the direction with a better channel condition, more source packets can be sent. The useful video rates can be expressed using  equations similar as \eqref{RV_12} and \eqref{RV_21}. We search over all candidate $R_{12}$, $R_{21}$, $R_{r}$ to choose the optimum parameter set that maximizes the average video rate.

To ensure all schemes use the same amount of total power, we use the normalized power consumption at relays such that if {\it K} nodes act as relays, then 1/{\it K} power is assigned at each node.  Due to the random nature of fading, the exact number of relay nodes at each fading realization of the network is unknown. To avoid information interchange between nodes and relays, we compute the average number of relays based on simulations for each ($R_{12}$ and $R_{21}$) for each scheme (R-DSTC and R-DSTC+NC) and use this number to normalize the power.

Nodes who can receive packets from user 1 can be around the user in all directions. Those nodes who are far away from user 2 are not likely to transmit correctly to user 2. Assigning a fraction of power to them is not energy-efficient, especially when the nodes are densely distributed. Therefore, we re-arrange the transmission order for Scheme II such that two users transmit in the first two hops while only those nodes who received packets from both directions relay the packets to their destinations separately in the last two hops. 
We refer this scheme with hop reordering as ``R-DSTC-HR".

\vspace{-1pt}	

\section{Simulation Results}\label{sec:simulation}

We use simulations to obtain the end-to-end PER under different scenarios. The bit error rate (BER) of a single link is first acquired through numerical simulations \cite{alay_RDSTC10}, for given transmission rate and fading level. The PER is then calculated following \cite{alay_RDSTC10}. Given a fading level, the convolutionally coded bits of a packet are randomly flipped according to the channel BER and the reception of this  packet is determined after channel decoding. We assume the fading level is constant over each packet duration, and it changes independently between packets.  For each scheme, we simulate transmission of 1000 packets with independent fading levels for each node placement, and then determine the average PER.

In the simulations, we assume two users are placed among a set of randomly uniformly distributed nodes with two different node densities, each of which having 15 different node distributions. For each node density, simulations are done at several user distances. We assume $P_{size}=1500$ Bytes, $\beta=0.1$ and set $T_{delay}$ to 150 ms in our simulations. We study an IEEE 802.11g based network and choose transmission rates from [6 9 12 18 24 36 48 54] Mbps for both user transmission and relay transmission. We only consider  transmission rates that lead to end-to-end PER less or equal to 25\%, as a link becomes unreliable when the PER $>$ 25\% \cite{alay_RDSTC10}.


For the FEC computations, we choose the number of parity packets such that the probability of packet level FEC decoding failure is below $\tau=0.5\%$. This threshold is chosen based on the observation that when using an error-resilient video decoder, there is no observable quality degradation \cite{alay_RDSTC10}.


We search all possible combination of parameters (transmission rates and time partition) for each node placement to get their corresponding end-to-end PERs and find the parameter set yielding maximal average useful video rate. This rate is further averaged over different node placements under the same node density.

\begin{figure}
	\centerline{\includegraphics[width=0.4\textwidth]{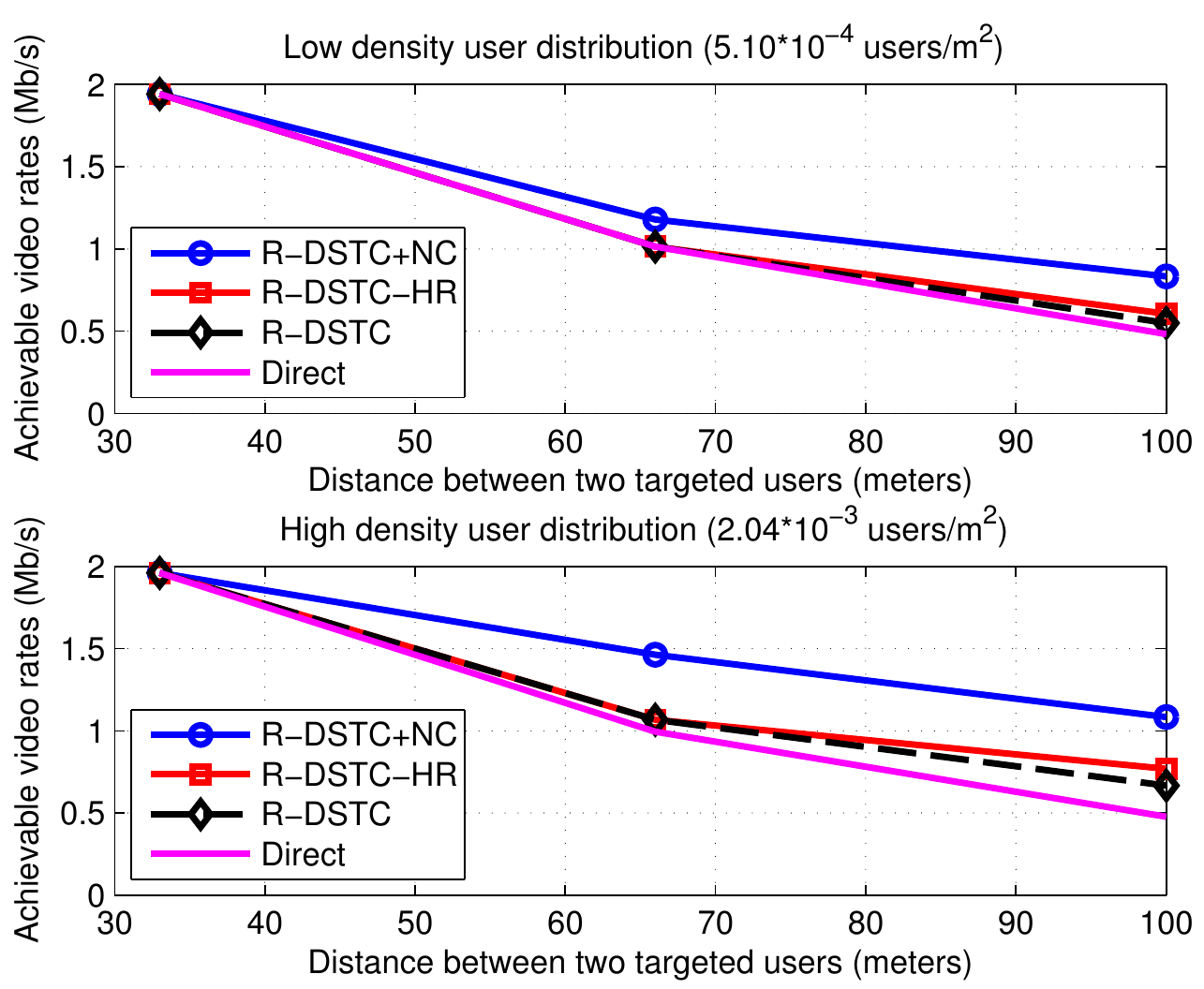}}
		\vspace{-5pt}	
	\caption{
		Useful video rates for 4 different schemes
	}
	\vspace{-22pt}	
	\label{fig: results}
\end{figure}

Fig. \ref{fig: results} compares the achievable average video rates by different schemes under different inter-user distances, for two assumed node densities. In Scheme II and III we allow the time slot for relay transmission to be 0, which reduces the system operation to direct transmission. We can see that the R-DSTC provides gain over direct transmission, with more gains at longer user distance and higher node density. It's also observed that R-DSTC-HR improves the rate over R-DSTC when the user distance is large by concentrating its power. Finally R-DSTC+NC provides additional significant gain over R-DSTC-HR.

A major difference between R-DSTC+NC and R-DSTC systems is that the number of hops is reduced from 4 to 3. If the transmission rates at all hops are the same, a performance gain of 4/3 is expected (provided that the end-to-end packet loss rates are also the same under these same transmission rates). However, the optimal parameter set yielding maximum video rate does not always use the same transmission rate at all hops. When the rates are unequal, the ratio of $n_{12}$ for Schemes III and II (see \eqref{eq_RDSTC} and \eqref{eq_NC}) can be different from 4/3. Results in Fig. \ref{fig: results} show that at user distance of 100 m, the gain of R-DSTC+NC over R-DSTC-HR is quite close to 4/3. At smaller distances, smaller gains are observed, whereas at larger distances, larger gains are observed.   

\begin{figure}
	\centerline{\includegraphics[width=0.45\textwidth]{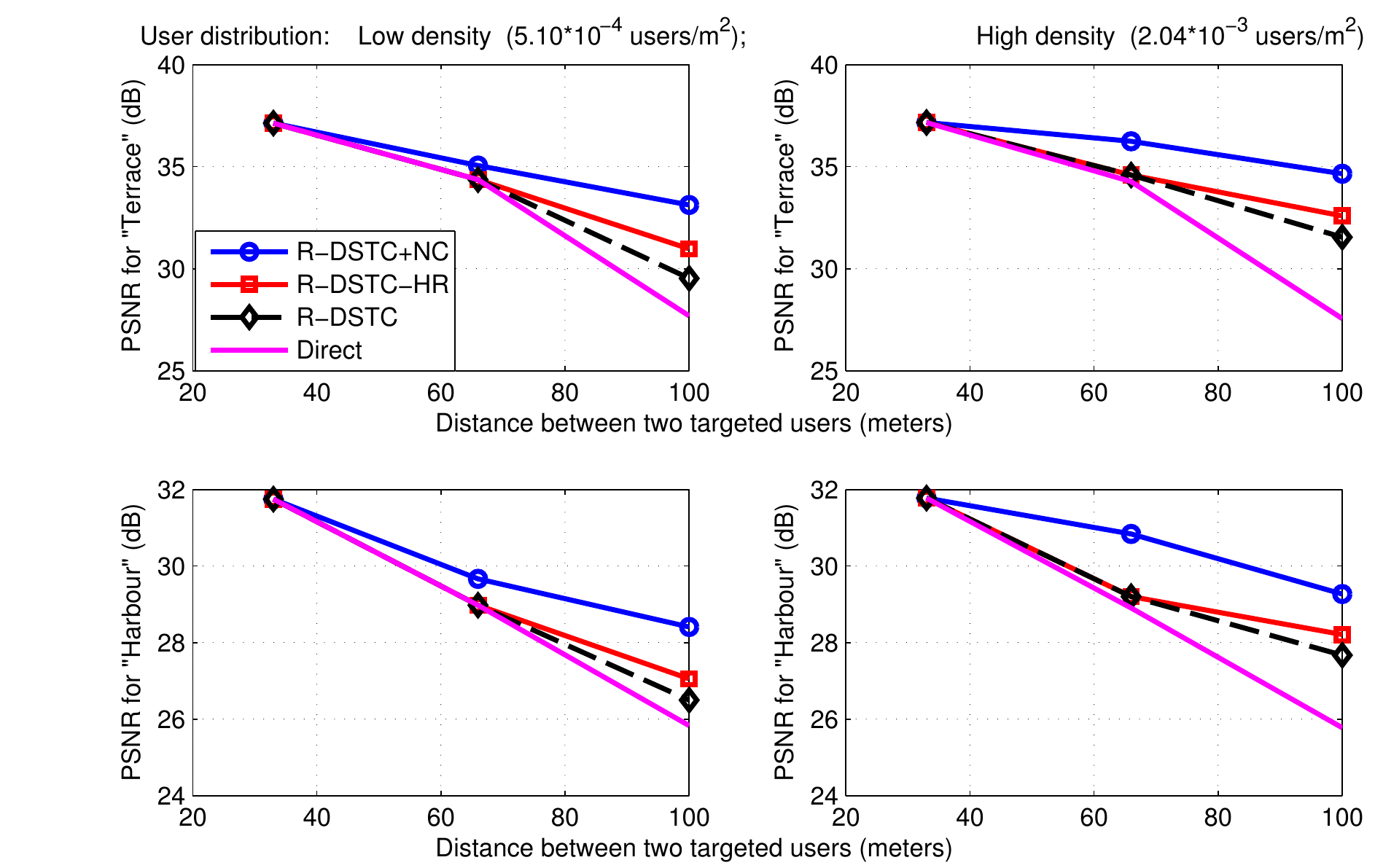}}
		\vspace{-5pt}	
	\caption{
		Video quality vs. user distances
	}
	\vspace{-22pt}	
	\label{fig: PSNR}
\end{figure}

To evaluate the video quality improvement due to the video rate increase, PSNR of decoded video versus user distance are plotted in Fig. \ref{fig: PSNR}. All the curves in Fig. \ref{fig: PSNR} are acquired by mapping the sustainable video rates into their corresponding PSNR values, which are obtained from the experimental video coding results using H.264/AVC reference software JM \cite{JM} for 704x576 video ``Terrace" and ``Harbour". We note that typically 1 dB gain in PSNR corresponds to noticeable visual improvements. From Fig. \ref{fig: PSNR}, as the user distance increases, video quality in all schemes decrease with schemes using relays having a slower rate of quality degradation. Note that at the longest distance considered, direct transmission gives PSNR at 26-27 dB, which corresponds to poor perceptual quality.  R-DSTC-HR provides considerable gain over Scheme I, and R-DSTC+NC provides additional 1-2 dB gain, leading to significantly better quality.

\vspace{-2pt}	
\section{Conclusion}\label{sec:conclusion}
\vspace{-2pt}	

In this paper, we propose a two-way video communication scheme using NC and R-DSTC along with packet level FEC to enable error resilient video transmission without feedback. We search the transmission rates and time partition under the given delay constraint, to maximize the average useful video (data) rate at both users. We show that the proposed scheme significantly outperforms both rate-adaptive direct transmission and using R-DSTC scheme only. We also observe that relaying is more beneficial when the node density is high and the distance between users is large.

In the future, we plan to extend this work to next generation cellular networks, such as 3GPP LTE and IEEE 802.16. In particular, we will used tractable  stochastic geometry models \cite{FutureWork} to capture the effect of  base station locations and user distributions in the network to the performance of randomized cooperation based video conferencing applications.


\end{document}